\documentclass{llncs}
\usepackage{graphicx}
\usepackage{amsmath}
\usepackage{amssymb}
\usepackage{amsfonts}
\usepackage[bookmarks]{hyperref}
\usepackage{multirow}
\usepackage{array}
\usepackage{float}
\usepackage{colortbl}
\usepackage{arydshln}
\usepackage{slashbox}
\usepackage{stmaryrd}
\usepackage{pifont}

\begin{document}

\title{Privacy-Preserving Contact Tracing: current solutions and open questions}
\author{Qiang Tang}

\institute{Luxembourg Institute of Science and Technology (LIST)\\
4362, Esch sur Alzette, Luxembourg\\
qiang.tang@list.lu
}

\maketitle


\begin{abstract}
The COVID-19 pandemic has posed a unique challenge for the world to find solutions, ranging from vaccines to ICT solutions to slow down the virus spreading. Due to the highly contagious nature of the virus, \emph{social distancing} is one fundamental measure which has already adopted by many countries. At the technical level, this prioritises \emph{contact tracing} solutions, which can alert the users who have been in close contact with the infected persons and meanwhile allow heath authorities to take proper actions.  In this paper, we examine several existing privacy-aware contact tracing solutions and analyse their (dis)advantages. At the end, we describe several major observations and outline an interdisciplinary research agenda towards more comprehensive and effective privacy-aware contact tracing solutions.
\end{abstract}

\section{Introduction}

The Coronavirus disease (COVID-19) pandemic, caused by the SARS-CoV-2 virus, has put the world into a panic mode. Up to now, the virus has contracted more than 2 million victims, and more than 120 thousands victims have lost their lives\footnote{\url{https://www.worldometers.info/coronavirus/}}. Among the survivors, many of them are in critical conditions and heavily rely on medical equipments such as ventilators to survive. It is reported that many victims have tragically died due to the shortage of such equipments\footnote{\url{https://en.wikipedia.org/wiki/2019-20_coronavirus_pandemic}}. While these numbers are still increasing on a daily basis, the world has united unprecedentedly to find solutions to suppress the pandemic.

According to the World Health Organization (WHO), on 31 December 2019, the WHO China Country Office was informed of a pneumonia of unknown cause, detected in the city of Wuhan in Hubei province, China. Soon, similar cases, later being attributed to COVID-19, appeared in Wuhan and Hubei province rapidly. To combat the epidemic, the Chinese government adopted a variety of extreme measures, e.g. completely close the borders of villages, cities, and provinces; track and then quarantine all close contacts of COVID-19 victims; make wearing masks mandatory. From the public information, the epidemic has been successfully controlled in China and only dozens of new cases are identified daily now. In a sharp contrast, the numbers are still rapidly growing in the western democracies, where some of these extreme measures cannot be easily enforced due to the potential violation of fundamental rights. Nevertheless, in order to slow down the virus spreading and reduce the pressure on medical systems, \emph{social distancing} is widely promoted and enforced, furthermore many countries are investigating the concept of \emph{contact tracing}. Without careful considerations, \emph{contact tracing} can turn into a massive surveillance tool so that individual's privacy can be seriously damaged, see the case analysis for China \footnote{\url{https://www.nytimes.com/2020/03/01/business/china-coronavirus-surveillance.html}} and for South Korea \footnote{\url{https://www.nature.com/articles/d41586-020-00740-y}}. What makes things harder is that the perception of privacy heavily depends on the political regime and the underlying culture, see the analysis by Asghar et al. towards Singapore's \emph{TraceTogether} app\footnote{\url{http://tiny.cc/fljqmz}}.

It remains an open problem to design privacy-aware \emph{contact tracing} solutions which also satisfy the requirements in other dimensions.

\subsection{SoTA of Privacy-aware \emph{contact tracing}}

Up to now, a number of \emph{contact tracing} solutions have been introduced. For instance, China and South Korea have begun tracing COVID-19 victims and their contacts from the very early stage of the epidemic, via smartphones as well as face recognition technologies. In addition, the Korean government even made a lot of collected data public\footnote{\url{https://coronamap.site/}}. On one hand, these mandatory tracing solutions greatly facilitate the containment of the virus, but on the other hand it also raises serious privacy concerns\footnote{\url{https://www.nature.com/articles/d41586-020-00740-y}} \cite{Ferretti20}.

Recently, the Singapore government  published an app, named \emph{TraceTogether} \footnote{\url{http://tiny.cc/onlqmz}}, which exploits some cryptographic primitives for privacy protection. This app has attracted the attention from some other countries, such as Australia, which are currently evaluating its privacy guarantee  according to their own privacy regulations, see \cite{aus2020}. At the end of March 2020, Israel passed an emergency law to launch a smartphone app to reveal if a user was, over the previous 14 days, in close proximity to anyone who has contracted the virus. Besides the efforts at the national level, initiatives have also been started by the general public. For instance, a Pan-European Privacy-Preserving Proximity Tracing
(PEPP-PT) project has been kicked off with both public and private partners from several EU countries\footnote{\url{https://www.pepp-pt.org/}}, MIT is collaborating with WHO to advocate its app named \emph{Private Kit: Safe Paths}\footnote{\url{http://safepaths.mit.edu/}}, Google and Apple are also collaborating on new solutions\footnote{\url{https://www.apple.com/covid19/contacttracing/}}. There are also more theoretical proposals, e.g. those from \cite{icc18,Brack2020,dp3t2020,iacr2020}.

In the meantime, the concept of \emph{contact tracing} and proposed solutions have been scrutinized by commenters and researchers. For instance, Anderson provide very insightful analysis on the practical aspects of \emph{contact tracing}\footnote{{\url{http://tiny.cc/2z0zmz}}}.
Wang gave some interesting remarks on current \emph{contact tracing} solutions\footnote{{http://tiny.cc/tx0zmz}}, including that from Google and Apple  solutions\footnote{\url{https://www.apple.com/covid19/contacttracing/}}. Asghar et al. analysed Singapore's \emph{TraceTogether} app\footnote{\url{http://tiny.cc/fljqmz}}, and Vaudenay \cite{Vaudenay2020} provided a detailed analysis of the DP-3T solution by Troncoso et al. \cite{dp3t2020}.

\subsection{Contribution}

In this paper, we aim at a deeper understanding about the utility and privacy issues associated with emerging \emph{contact tracing} solutions, particularly those related to respiratory system diseases such as COVID-19. Our contribution lies in three aspects.

\begin{enumerate}
\item We analyse the application context in this  COVID-19 pandemic and identify a broad set of utility and security requirements.

\vspace{0.05cm}

\item We examine several existing privacy-aware \emph{contact tracing} solutions and analyse their (dis)advantages. These solutions include the \emph{TraceTogether} app from Singapore\footnote{\url{http://tiny.cc/onlqmz}} and three cryptographic solutions by Reichert et al. \cite{iacr2020} and Altuwaiyan et al. \cite{icc18} and Troncoso et al. \cite{dp3t2020}. It is worth noting that many similar solutions exist and more apps are being developed, and we wish our analysis can be extended to them.

\vspace{0.05cm}

\item We summarize our findings into several major observations and outline an interdisciplinary research agenda towards more comprehensive and effective privacy-aware \emph{contact tracing} solutions.
\end{enumerate}

Note that many of the above works have not been formally published, so that they might be updated by their authors from time to time. Nevertheless, we base our analysis on some specific versions which were available when this paper has been written, as indicated in the references.

\section{A Closer Look at Contact Tracing}

In the public health domain, \emph{contact tracing} refers to the process of identification of contacts who may have come into contact with an infected victim and subsequent collection of further information about these contacts. By tracing the contacts of infected individuals, testing them for infection, treating the infected and tracing their contacts in turn, public health aims to reduce infections in the population\footnote{\url{https://en.wikipedia.org/wiki/Contact_tracing}}. In practice, \emph{contact tracing} is widely performed for diseases like sexually transmitted infections (including HIV) and virus infections (e.g. SARS-CoV and SARS-CoV-2/COVID-19). Despite some pioneering attempts in applying advanced ICT technologies, e.g. the \emph{FluPhone} \cite{Yoneki11}, \emph{contact tracing} has mainly been implemented manually by medical personnel. Regardless, this approach has been proven effective in combating contiguous diseases because it can at least (1) interrupt ongoing transmission and reduce spread, alert contacts to the possibility of infection and offer preventive counseling or prophylactic care and (2) allow the medical professionals to learn about the epidemiology in a particular population.

\begin{figure}[h]
\centering
\includegraphics[scale=0.18]{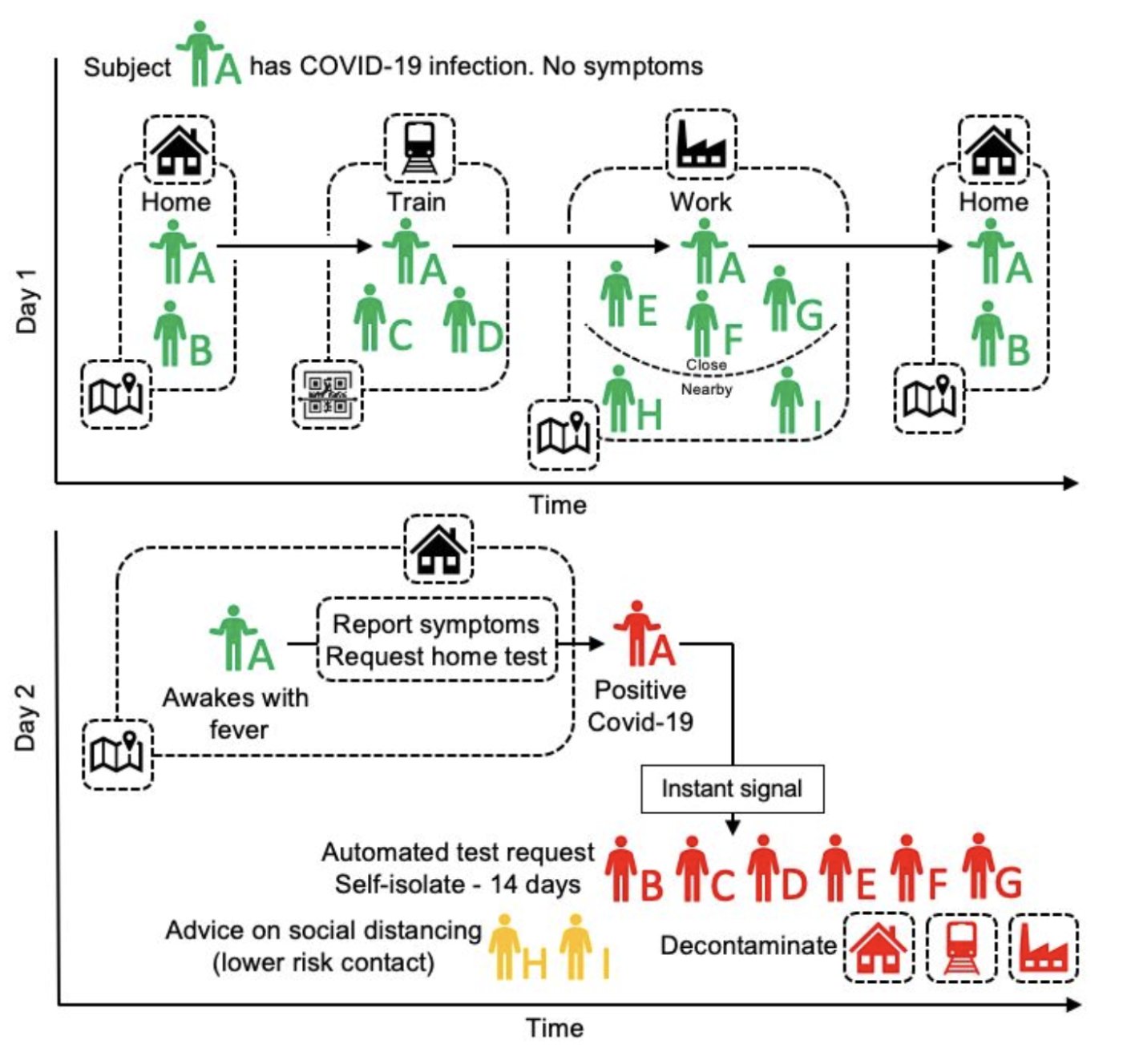}
\caption{Quarantine Plan (taken from \cite{Ferretti20})}
\label{quarantine}
\end{figure}

Ferretti et al. \cite{Ferretti20} investigated the key parameters of epidemic spread for COVID-19 and concluded that viral spread is too fast to be contained by manual \emph{contact tracing} (see an illustration in Fig. \ref{quarantine}) but could be controlled if this process was faster, more efficient and happened at scale. This necessitates digital \emph{contact tracing} solutions, such as those based on mobile apps, which can efficiently achieve epidemic control in large scale. Interestingly, this coincides with the empirical analysis of the control measures in China, by Tian et al. \cite{Tian20}.

In the following-up discussions, we focus on digital \emph{contact tracing} solutions, by cautiously referring to their manual ancestors in the aspect of functional requirements. For simplicity, we omit the term ``digital" in the rest of the paper.

\subsection{Preliminary on \emph{contact tracing}}

Overall, the purpose of a \emph{contact tracing} solution is to prevent an epidemic or a pandemic caused by a contagious virus or something similar. Depending on the standing point, it may serve (at least) two purposes.
\begin{itemize}
\item At a global level, it helps medical personnel to trace the pattern of virus spreading, produce transmission graphs, trace the origin of the virus, and so on. With adequate knowledge of the virus, the authority can take appropriate actions (e.g. disinfecting a facility) and make appropriate plans (e.g. enforcing \emph{social distancing}) to fight against the virus and prevent future similar epidemic or pandemic.

\vspace{0.05cm}

\item At an individual level, it helps the medical personnel to alert individuals who might have been infected. Alternatively, it may allow an individual to evaluate his/her risks of being infected and take further actions. As a quick remark, most cryptographic solutions focus on the individual-level \emph{contact tracing} and devoted their attention to privacy protection for individuals.
\end{itemize}

For many diseases, such as COVID-19, an individual might contract the virus with either direct or indirect contact to the infected person. In case of \emph{direct contact}, we can imagine that the droplets containing virus from the infected can fly to his mouth, nose, or eyes. In addition, the droplets might also attach to his clothes. In case of \emph{indirect contact}, we can imagine that the infected leaves virus-droplets on a book, a chair, or any physical objects, and later on an individual might tough the object and contract the virus. Taking the COVID-19 as an example, the virus can be transmitted through either direct or indirect contacts, so that it makes comprehensive \emph{contact tracing} a very hard problem.

To design a \emph{contact tracing} solution, the main anchor is location data. Intuitively, if two persons are located in close locations at a certain point of time then we can informally assume that they have close contact with each other\footnote{Note that there are exceptions though.}. In reality, other anchoring technologies might be employed, such as face recognition and other AI-based tracking technologies. However, we do not consider them because such technologies are only deployed in very limited regions of the world. With the abundance of electronic gadgets, location data can be generated and collected in many ways, e.g. GPS, WIFI, Telcom Cell Towers, Bluetooth beacons. Broadly, we can categorize location data into two categories.
\begin{itemize}
\item One is \emph{absolute location} data. In this category, we can think of GPS location, location with respect to static WIFI access points and Telcom cell towers. A location data point can often be written in the form of geolocation coordinate pair.

\item The other is \emph{relative location} data. In this category, we can think about the pairing of two Bluetooth-enabled smart devices, the boarding on a transportation tool such as planes, buses, cars, or ships. In this case, we can have some reference description about the location, for example both persons are on the same flight on the day XYZ.
\end{itemize}
Besides the difference in collection and management, when being applied in \emph{contact tracing} solutions, they also have very different precision and security implications.

With respect to precision, \emph{absolute location} data is often generated by external infrastructures and might not be precise enough to define ``close contact" in a epidemic and pandemic. On the other hand, such location data can be collected constantly and will provide a big picture on the mobility patterns of individuals. In contrast, \emph{relative location} could be more precise. But in order to collect such data, we need to assume a large potion of the population will install the same app to support the service. Another drawback of this type of location data is that it might be ad hoc and will not be able to provide a comprehensive view of individuals' mobility history. Moreover, it is hard to use such data alone to study the transmission pattern of a disease in a population. It seems that, in order to facilitate the global and individual level objectives, a \emph{contact tracing} solution should be based on fusing location data of both categories.

With respect to security, two aspects are of crucial importance. One aspect is data authenticity. In some scenarios, \emph{absolute location} data could be more authentic because a third party could offer some sort of attestation. For instance, a Telcom operator can attest the location of an individual's smartphone. In contrast, it is harder to find attestations on an individual's \emph{relative location} data. However, there are exceptions. For instance, if a user presented a boarding pass for a flight, then it can be an attestation that this user is on the plane during a certain period of time. The other aspect is privacy. Naturally, we can imagine that authentic data has tight bound to the identity of an individual so that it will be more privacy invasive if disclosed. It is a difficult task to balance the authenticity and privacy of location data and some tradeoff might be inevitable.


\subsection{System Architecture and Requirements}

\begin{figure}[h]
\centering
\includegraphics[scale=0.46]{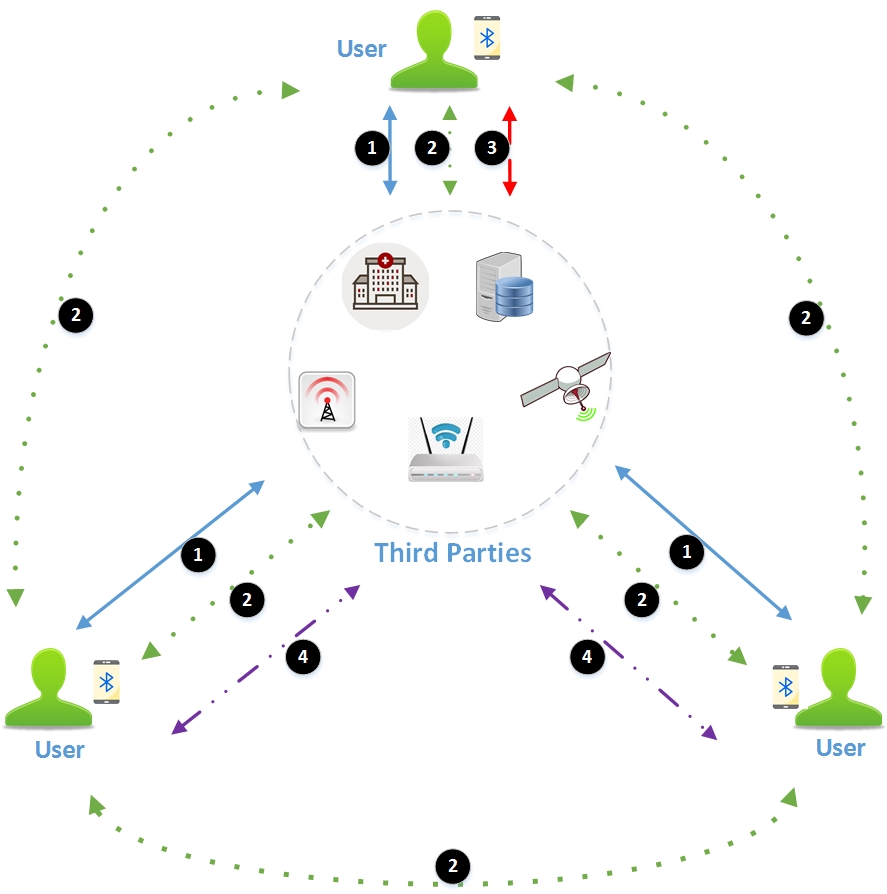}
\caption{System Architecture}
\label{architecture}
\end{figure}

With respect to the existing \emph{contact tracing} solutions, there is neither a uniform system architecture nor a defined set of participants. Nonetheless, all the potential participants can be divided into two groups. In one group, there are individual users, who either have been confirmed with infection or have the risk of being infected when in close contact with the infected. In the other group, there are third parties, which vary in specific solutions. For instance, one potential player in this group is the health authority and medical personnel, who need to evaluate the situation and help the individual users if necessary. If a \emph{contact tracing} solution relies on smartphone apps, then the developer of this app may also be involved. In case that individual users needs to communicate with each other, then a server may be required to facilitate the communication. Without loss of generality, the system architecture can be illustrated as in Fig. \ref{architecture}, where (\ding{202}, \ding{203}, \ding{204},\ding{205}) indicate the four phases in the workflow of a \emph{contact tracing} solution, i.e. (\emph{initialisation}, \emph{sensing}, \emph{reporting}, \emph{tracing}).

\begin{itemize}
\item In the \emph{initialisation} phase, individual users and relevant third parties need to set up the system to enable the operations in other phases. For example, every individual user might be required to have a smart phone and download an app from a third party. In addition, cryptographic credentials may need to be generated and distributed.

\vspace{0.05cm}

\item In the \emph{sensing} phase, individual users will record their own location trails and also collect location data from their close contacts.

\vspace{0.05cm}

\item In the \emph{reporting} phase, if an individual user is confirmed to be infected then she needs to collaborate with some third parties to make her relevant location data available for the further uses.

\vspace{0.05cm}

\item In the \emph{tracing} phase, some third parties could collect and aggregate the location data from infected individuals for any possible legitimate purposes. For instance, a third party can communicate with the close contacts of the infected, or let the uninfected individuals to evaluate the risks of being infected on their own.

\end{itemize}

In comparison to other digital solutions (e.g. a general-purpose social app), a \emph{contact tracing} solution is expected to provide stricter guarantees towards utility and security.

As to utility, the solution should provide fine-grained and accurate measurement of the distance between a contact and the infected victim. The ``fine-grained" requirement refers to a precise description of the occurring time and the duration of the contacting event, while the ``accurate" requirement means that the error in the distance measurement should be as small as possible. Take the COVID-19 as an example, it is commonly considered that there is a risk when a contact and the infected victim stay together within 2 meters. In this case, if a GPS system has the error around 5 meters, then it should not be used alone in any solution.

The security guarantee imposes requirements on the protection for both authenticity and privacy.
\begin{itemize}
\item The ``authenticity" requirement means that the reported location data from a user must be real and should not have been forged or modified by anybody including the user herself. Lacking of authenticity can cause a lot of serious issues, as happened in China and South Korea. For example, an infected victim can blackmail a shop by claiming that she has been a visitor, an infected victim can cause a social panic by claiming that she has visited some heavily populated areas such as shopping malls or train stations, a user can forge location data in order to probe the location data of infected victims and identify them given that some matching service is offered, and fake location data from infected users can mislead medical personnel in their professional activities.

    In addition, the ``authenticity" requirement can be extended to the binding property between the location data to an individual. Quite often, smart devices are used to collect and manage location data, while such devices can be shared by several individuals. Lacking of binding can lead to some fraud activities. For example, an individual user can present another user's device and location data to get a priority in virus testing or avoid going to work by triggering some quarantine policy on purpose.

\vspace{0.05cm}

\item The ``privacy" requirement applies to both infected victims and other users. Except for disclosing the necessary information to the authorities, an infected victim might want to prevent any further disclosure to avoid social embarrassment or discrimination. An individual user may want to check his risk of being infected by matching his location data with that from infected, but he may not want to disclose his data to the third parties or the infected for any other purpose.

\end{itemize}

Note that the DP-3T consortium recently published a summary of privacy and security threats\footnote{ \url{https://github.com/DP-3T/documents}}.

\section{Analysis of Existing Solutions}

In this section, we provide some high-level analysis of several solutions against our formulation in Section 2. The common issues with these solutions are summarized into the major observations in Section 4.

\subsection{Singapore's \emph{TraceTogether}}

The \emph{TraceTogether} protocol from Singapore, as recapped in \cite{aus2020}, has two types of entities, namely users ($1 \leq i \leq N$) and the Ministry of Health (MoH) of the Singapore government. It is assumed that all users trust the MoH to protect their information. Note that, as shown below, the users are not required to share everything with MoH if they have not been in close contact with any confirmed COVID-19 victim. The protocol is elaborated below.

\begin{itemize}
\item In the \emph{initialisation} phase, a user $i$ downloads the \emph{TraceTogether} app and install it on her smartphone. The app sends the phone number $NUM_i$ to MoH and receives a pseudonym $ID_i$. MoH stores the ($NUM_i, ID_i$) pair in its database.

    MoH generates a secret key $K$ and selects an encryption algorithm $\mathsf{Enc}$. At the beginning of the app launch, MoH decides some time intervals $[t_0, t_1, \cdots]$, which will end when the pandemic is over. For the user $i$, MoH pushes $TID_{i,x} = \mathsf{Enc}(ID_i, t_x; K)$ to user $i$'s app at the beginning of $t_x$, for $x \geq 0$.

\vspace{0.05cm}

\item In the \emph{sensing} phase, user $i$ broadcasts $TID_{i,x}$ at the time interval $[t_x, t_{x+1})$ for all $x \geq 0$. For example, when user $i$ and user $j$ come into a range of Bluetooth communication at the interval $[t_x, t_{x+1})$, then they will exchange $TID_{i,x}$ and $TID_{j,x}$. They will store a $(TID_{i,x}, TID_{j,x}, Sigstren)$ locally in their smartphones, respectively. The parameter $Sigstren$ indicates the Bluetooth signal strength between their devices.

\vspace{0.05cm}

\item In the \emph{reporting} phase, suppose that user $i$ has been tested positive for COVID-19, then she is obliged to share with MoH the locally-stored pairs $(TID_{i,x}, TID_{j,x})$ for all relevant $j$ and $x$.

\vspace{0.05cm}

\item In the \emph{tracing} phase, after receiving the pairs from user $i$, MoH decrypts every $TID_{j,x}$ and obtains $ID_j$. Based on $ID_j$, MoH can looks up $NUM_j$ and then contact user $j$ for further instructions.

\end{itemize}

Whether or not to install the \emph{TraceTogether} app is a voluntary choice for the Singapore residents, and it is not clear how many installations have been made until this moment. So far, we have not found any official information showing how much this app has contributed to Singapore's war against COVID-19. Soon after its launch, Asghar et al. pointed out some privacy concerns\footnote{\url{http://tiny.cc/fljqmz}}. We have the following additional comments.

With \emph{TraceTogether}, individual users are required to put more trust on the third party - MoH, than in other solutions. Note that this might be a result of the political regime and cultural status of Singapore. To prevent a curious MoH from learning unnecessary mobility information, the mobility data of low-risk individuals are not required to be uploaded to MoH. However, things will change when more and more individual users are infected and diagnosed. By then, MoH will have decrypted data for most of the individuals and learned their \emph{relative location} data at certain points of time. If a big portion of the Singapore population has deployed the app, and then the location data on their smartphones would provide MoH a very clear mobility view of the whole Singapore population.

Without using secure hardware or other trusted computing technology, a malicious user can potentially manipulate the location data collected by the app, e.g. delete or add entries. Moreover, an attacker can mount relay attacks, e.g. to relay the Bluetooth signal from Alice's smartphone to Bob's smartphone even when they are far from each other. To this end, some other attacks, demonstrated in Vaudenay's analysis \cite{Vaudenay2020} against DP-3T solution by Troncoso et al. \cite{dp3t2020}, can also apply. Regarding the solutions employing Bluetooth for distance measurement, one attack deserves special attention. An attacker can place a \emph{Bluetooth range extender} in a relatively populated place, such as a city square, and as a result it will make any pair of users a close contact to each other. Such an attack is easy to mount and will seriously distort the functionality of the underlying solution.

In comparison to  other solutions, one advantage of \emph{TraceTogether} is that it offers MoH the ability to draw the transmission graph of COVID-19 in the population which has installed the app. It can be easily done based on the fact that the temporary identifiers are linked to the phone numbers, which can help MoH identify the individual users. This advantage is an outcome of the tradeoff between privacy and utility.

\subsection{Reichert et al.'s MPC Solution}

In the solution proposed by Reichert et al. \cite{iacr2020}, it is assumed that user $i$ $(1 \leq i \leq N)$ possesses a smart device that can collect and store geolocation data. In addition, a Health Authority (HA) will collect the geolocation history of all COVID-19 victims and offers data matching as a service.

\begin{itemize}
\item In the \emph{initialisation} phase, HA prepares the cryptographic key materials for generating garbled circuits for later use.

\vspace{0.05cm}

\item In the \emph{sensing} phase, user $i$ records her geolocation data points on the fly. Let the time intervals be denoted by $[t_0, t_1, \cdots]$. At time $t_x$, user $i$ generates and stores a tuple $(t_x, l_{x,u}, l_{x,v})$, where $l_{x,u}$ and $l_{x,v}$ represent the latitude and longitude of the location, respectively.

\vspace{0.05cm}

\item In the \emph{reporting} phase, if user $i$ has been tested positive for COVID-19, then she shares with HA her $(t_x, l_{x,u}, l_{x,v})$ for all relevant $x$.

\vspace{0.05cm}

\item In the \emph{tracing} phase, suppose user $j$ wants to check whether he has been in close contact with any COVID-19 victim. For any COVID-19 victim user $i$, HA constructs a garbled circuit based on her geolocation data points and shares the circuit with user $j$, who can then retrieve some key materials from the HA and privately evaluate the circuit based on his own geolocation data points. Note that user $j$ is assumed to have high risk if, for some of her data point(s), both the time stamps and the geolocation coordinates are close to one of that from user $i$.

\end{itemize}

This is a theoretical cryptographic solution, which covers the matching between an infected and other individual users while ignoring other aspects of a \emph{contact tracing} solution. With respect to the cryptographic design, scalability will be a bottleneck. For any user $j$, HA will need to prepare garbled circuits for all the infected victims, and interact with user $j$ in the \emph{tracing} phase. When the sizes of infected populations and the requesters become large, the complexity for HA will be formidable. In addition, malicious users might leverage this to mount (D)DoS attacks, unless proper countermeasures are deployed. Another issue is the lacking of details on the computation of infection risks, which are based on the proximity of both stamps and the geolocation coordinates. It is unclear how to set a threshold on the time stamps, considering there are a variety of mobility patterns between the concerned users.

\subsection{Altuwaiyan et al.'s Matching Solution}

In the solution proposed by Altuwaiyan et al. \cite{icc18}, it is assumed that user $i$ $(1 \leq i \leq N)$ possesses a smart device that can exchange information (e.g. Bluetooth messages) with similar devices nearby. In addition, a server will collect the data of all infected victims and offer data matching as a service.

\begin{itemize}
\item In the \emph{initialisation} phase, there is no special setup for the server and users.

\vspace{0.05cm}

\item In the \emph{sensing} phase, user $i$ exchanges information with similar devices nearby. Let the time intervals be denoted by $[t_0, t_1, \cdots]$. At time $t_x$, user $i$ generates and stores a tuple of data points $(t_x, (m_{i,1}, r_{i,1}, p_{i,1}), \cdots, (m_{i,n_{i,x}}, r_{i,n_{i,x}}, p_{i,n_{i,x}}))$, where $(m_{i,1}, r_{i,1}, p_{i,1})$ records information about the first encountered device, where $m_{i,1}$ is the hashed identifier of this device, $r_{i,1}$ is the detected signal strength and $p_{i,1}$ is the type of device,  and so on.

\vspace{0.05cm}

\item In the \emph{reporting} phase, if user $i$ has been tested positive, then she shares with the server her data $(t_x, (m_{i,1}, r_{i,1}, p_{i,1}), \cdots, (m_{i,n_{i,x}}, r_{i,n_{i,x}}, p_{i,n_{i,x}})$ for all relevant $x$.

\vspace{0.05cm}

\item In the \emph{tracing} phase, suppose user $j$ wants to check whether he has been in close contact with some infected victim, he generates a public/private key pair $(pk,sk)$ for a homomorphic encryption scheme. Then, user $j$ sends the timestamps in his data to the server, which will find all the infected users who have some overlapped timestamps. For each such infected user $i$ and overlapped timestamp $t_x$, the server and user $j$ perform the following protocol.
    \begin{enumerate}
    \item User $j$ sends $(\mathsf{Enc}(m_{j,1}, pk), \cdots, \mathsf{Enc}(m_{j,n_{j,x}}, pk))$ to the server.
    \item The server computes the following matrix and sends it to user $j$.

        \begin{table}[h]
        \centering
        \begin{tabular}{|p{4.6cm}|p{1cm}|p{4.8cm}|}\hline

        $\mathsf{Rand}(\mathsf{Enc}(m_{j,1}-m_{i,1}, pk))$ &  $\cdots$  & $\mathsf{Rand}(\mathsf{Enc}(m_{j,n_{j,x}}-m_{i,1}, pk))$ \\    \hline

        $\vdots$ &  $\vdots$  & $\vdots$ \\    \hline

        $\mathsf{Rand}(\mathsf{Enc}(m_{j,1}-m_{i,n_{i,x}}, pk))$ &  $\cdots$  & $\mathsf{Rand}(\mathsf{Enc}(m_{j,n_{j,x}}-m_{i,n_{j,x}}, pk))$ \\    \hline
        \end{tabular}
        \end{table}

        In the table, $\mathsf{Rand}$ is a ciphertext randomization function. If the ciphertext encrypts 0 then the randomized ciphertext still encrypts 0, otherwise the randomized ciphertext will encrypt a random number.
    \item User $j$ decrypts every ciphertext in the received matrix, and obtains the $m_{j,y}$, where $1 \leq y \leq n_{j,x}$, which is overlapped with that from user $i$. Then, user $j$ sends all the matched $m_{j,y}$ together with $r_{j,y}$ to the server.

    \item Based on the data from user $j$, the server computes the distance between user $i$ and user $j$, and then acts accordingly.
    \end{enumerate}

\end{itemize}

We note that, in addition to the aforementioned privacy-aware matching protocol, the other contribution of Altuwaiyan et al. \cite{icc18} is a new method to measure the distance between two smart devices. In comparison to other solutions where the distance is measured based on the perceived signal strength of the peer device, the method leverages on the signal strengths of more devices so that it is more accurate in practice.

Regarding privacy, the infected victims reveal the location data to the server, where the location data contain hashed network identifiers such as those of WIFI access points. Give that network identifiers are often static, this allows the server to recover the \emph{absolute location} data points of the infected users. If a match has been found in the \emph{tracing} phase, then user $j$'s  \emph{absolute location} at some time stamps are also revealed to the server. In addition, in order to improve computational efficiency, time stamps are always disclosed to the server. The revelation of time stamps and \emph{absolute location} implies serious privacy leakage, and should be avoided.

\subsection{DP-3T Solution}

In the solution proposed by Troncoso et al. \cite{dp3t2020}, it is assumed that user $i$ $(1 \leq i \leq N)$ possesses a smart device that can collect and store data. In addition, there is a backend server, and the Health Authority (HA). Note that the backend server acts as a communication platform to facilitate the matching activities among the users. Let $\mathsf{H}$, $\mathsf{PRG}$ and $\mathsf{PRF}$ denote a cryptographic hash function, a pseudorandom number generator and a pseudorandom function, respectively.

\begin{itemize}
\item In the \emph{initialisation} phase, user $i$ generates a random initial daily key $SK_{i,0}$, and computes the following-up daily keys based on a chain of hashes: i.e. the key for day 1 is $SK_{i,1}=\mathsf{H}(SK_{i,0})$ and the key for day $x$ is $SK_{i,x}=\mathsf{H}(SK_{i,x-1})$. Suppose $n$ ephemeral identifiers are required in one day, then the identifiers for user $i$ on the day $x$ are generated as follows:
    \begin{displaymath}
    EphID_{i, x, 1}||\cdots||EphID_{i, x, n}= \mathsf{PRG}(\mathsf{PRF}(SK_{i,x}, ``broadcast key"))
    \end{displaymath}

\vspace{0.05cm}

\item In the \emph{sensing} phase, on the day $x$, user $i$ broadcasts the ephemeral identifiers $\{EphID_{i, x, 1}, \cdots, EphID_{i, x, n}\}$ in a random order. At the same time, her smart device stores the received ephemeral identifiers together with the corresponding proximity (based on signal strength), duration, and other auxiliary data, and a coarse time indication (e.g., ``The morning of April 2”).

\vspace{0.05cm}

\item In the \emph{reporting} phase, if user $i$ has been tested positive for COVID-19, then HA will instruct her to send $SK_{i,x}$ to the backend server, where $x$ is the first day that user $i$ becomes infectious.

    After sending the $SK_{i,x}$ to the backend server, user $i$ chooses a new daily key $SK_{i,y}$ depending on the day when this event occurs. While not mentioned in \cite{dp3t2020}, we believe this new key should also be sent to the backend server, as user $i$ might continue to be infectious.

\vspace{0.05cm}

\item In the \emph{tracing} phase, periodically, the backend server broadcasts $SK_{i,x}$ after user $i$ has been confirmed with the infection. On receiving $SK_{i,x}$, user $j$ can recompute the ephemeral identifiers for day $x$ as follows

    \begin{displaymath}
    \mathsf{PRG}(\mathsf{PRF}(SK_{i,x}, ``broadcast key")).
    \end{displaymath}
    Similarly, user $j$ can compute the identifiers for day $x+1$ and so on. With the ephemeral identifiers, user $j$ can check whether any of the computed identifiers appears in her local storage. Based on the associated information, namely ``proximity, duration, and other auxiliary data, and a coarse time indication", user $j$ can act accordingly.

\end{itemize}

Vaudenay \cite{Vaudenay2020} provided detailed security and privacy analysis against this solution. It argues that decentralisation introduces new attack vectors against privacy, contrasting to the common belief that decentralisation helps solve the privacy concerns in centralised systems. One of the conclusions from \cite{Vaudenay2020} is that trusted computing technology seems unavoidable in order to prevent all the identified attacks.

As noted by Vaudenay \cite{Vaudenay2020}, a lot of practical details are missing from the whitepaper \cite{dp3t2020}. For instance, it is unclear how to instantiate the backend server. With regard to its pan European ambition, it is not clear how HAs and backend servers from different countries can efficiently coordinate with each other to take prompt responses. It also remains open how users' privacy will be infected if the credentials of some HAs and backend servers are compromised. We want to point out that the hashchain-based method of generating daily keys, see more information in the \emph{initialisation} phase, brings linkability risks when an individual's credential is leaked to an attacker, e.g. via malware. For example, if the attacker learns $SK_{i,x}$ then it can compute identifiers for day $x+1$ and so on. Then, the attacker can link user $i$ to his broadcasted identifiers, either collected by the attacker itself or or bought from other attackers. This could lead to disclosure of \emph{absolute location} data points of user $i$.

By design, the DP-3T protocol favors more the privacy for user $j$ than the infected user $i$. Referring to the \emph{tracing} phase description, if an identifier $EphID^*$ appears in his local storage user $j$ learns the exact time stamp when the perceived risk occurs. This can very likely enable user $j$ to link $EphID^*$ to the real person, whose device has broadcasted $EphID^*$, due to the fact that user $j$ can be close to very few users at a specific time stamp. Moreover, this also raises a concern of \emph{targeted identification attack}. If an attacker wants to find out whether or not some users have been infected, then he can simply get close enough to them and make their devices exchange identifiers (of course he should record the time stamps), and finally he can make a decision if identifiers have been matched at these time stamps. In contrast, the \emph{TraceTogether} avoids this problem by letting MoH calculate the infection risk for encountered users. Overall, the DP-3T protocol leaks too much unnecessary information about the infected users to the public and raises serious privacy concerns without deploying any further countermeasure. One possible solution is to deploy a two party secure computation protocol between the backend server and user $j$, for the latter to only learn a risk score. How to scale this up is a very practical problem to be addressed.

\section{Discussion and a Future Research Roadmap}

After examining many privacy-aware \emph{contact tracing} solutions, we feel that the situation is a bit chaotic at this moment. There is a rush for the academia to propose new technical solutions and for the industry to launch new apps, and new initiatives pop up every day. These efforts will somehow contribute to the fight against the pandemic, at least raise the awareness and help start discussions. Unfortunately, it is a pity that many solutions are based on assumptions which are unrealistic and hamper their usefulness and massive adoption. In this section, we first discuss our major observations, partly based on the analysis in Section 3,  and then put forward a roadmap which foresees an interdisciplinary research agenda for the future.

The first major observation is that many solutions only focused on the \emph{direct contact} scenario and paid little attention to the precision of location data and the accuracy of distance measurement (e.g. those based on Bluetooth signal strength). In addition, a number of exceptional situations have not been taken into account. For instance, the neighbours of an infected and quarantined victim can be in very close range and be classified to be of high risk, while the truth is that they are in different apartments or houses so that the risk of being infected is low. Inaccurate distance estimation and false risk alerts will cause unnecessary panic in the population and also waste a lot of resources in the healthcare system to address the fake suspects. Moreover, the scenario of \emph{indirect contact} has been ignored most of the time. Due to the fact that COVID-19 can be transmitted via indirect contact. Ignoring this scenario makes the existing \emph{contact tracing} solutions less practical and effective than expected.

The second major observation is that the scope of \emph{contact tracing} in existing solutions is very narrow. It is mostly about an individual evaluates his risk of being infected based on his contact history with the infected victims. However, as we have described in the beginning of Section 2 and Section 2.1, \emph{contact tracing}  is supposed to enable health authority and medical personnel not only to study the epidemic or pandemic at the global level but also to help the individual at the individual level. This raises an open question how the health authority and medical personnel can collect the necessary information and perform their normal duties. Inevitably, new privacy-aware solutions need to be designed and implemented to fulfill the objectives of a comprehensive \emph{contact tracing} yet minimizes information disclosure. One root problem, facing both existing and new solutions, is that a consensus on the trust relationship between the different players in the scope of \emph{contact tracing} is missing. When an app is deployed, more practical questions can occur. For example, how the results from the app of Google and Apple\footnote{\url{https://www.apple.com/covid19/contacttracing/}} should be interpreted, how the liability is distributed, and can an individual interact with the health authority and medical personnel on the basis of these results?

The third major observation is that most solutions only emphasized the privacy concerns in \emph{contact tracing} while paying no attention to the authenticity aspect (elaborated in Section 2.2). It implies that individual users can potentially forge location data for herself and for others as well.  Consequently, the results of \emph{contact tracing} could have been manipulated to a large extent, and enable the dishonest users and attackers to disrupt the service and commit various fraud activities. Of particular importance is that the lacking of authenticity can also lead to the breach of privacy, as demonstrated by Vaudenay \cite{Vaudenay2020}. How to balance authenticity and privacy seems to be the biggest challenge in designing a privacy-aware and effective \emph{contact tracing} solution.

The fourth major observation is that most theoretical solutions have not provided all the technical details to facilitate an implementation. In many solutions, an un-trusted backend server is required. It is unclear how this server can be chosen in practice and how to incentivize it to perform in the same way as what has specified. Furthermore, a health authority is often involved and required to perform some cryptographic operations. This seems to be an  unrealistic requirement for an governmental authority in many countries. At least, it will not be easily done in a short period of time.

To fight against a pandemic, like COVID-19, every technology and solution could matter. Nevertheless, we believe it is also important to set an interdisciplinary agenda to comprehensively formulate the problem, identify the requirements, and search for the opportunities. We foresee the following key research elements on the roadmap.

\begin{itemize}
\item Bridge the gap between health authority and solution designers. This essentially requires the solution designers to figure out what an effective \emph{contact tracing} solution needs to generate, for both individual users, health authority and the medical personnel. Accordingly, different players' roles should be clearly defined. Decisions should be drawn on the basis of regulations (at least) related to healthcare and privacy.

\vspace{0.05cm}

\item Evaluate and model the privacy and other security risks. This will need to clarify the trust relationships among the players and result in a set of security requirements, which should be satisfy by a solution. It should also reflect the accountability and liability configuration among the players. Various tradeoffs could be inevitable among privacy, utility, efficiency and other aspects.

\vspace{0.05cm}

\item Build incentive mechanisms into solutions from the beginning. Instead of simply providing a dichotomy choice through ``opt-in" and ``opt-out", it is important to incentivize the participation of individual users and other players, e.g. by employing technologies such as DLT or Blockchain. It is also important to deploy mechanisms to deter dishonest and malicious behaviours and encourage honest behaviour for the society good. In particular, it should prevent the solutions from being used in any manner as a surveillance tool for either political or economic purposes.
\end{itemize}

\section*{Acknowledgement}
This work is partially funded by the European Unions Horizon 2020 SPARTA project, under
grant agreement No 830892.

\bibliography{ConTrack}
\bibliographystyle{plain}

\end{document}